\documentclass[amsmath,amssymb,twocolumn]{revtex4}
\usepackage{graphicx}
\makeatletter \@addtoreset{equation}{section} \makeatother

\def\tanh{{\rm tanh}}

\def\Imag{{\rm Im}}

\def\ri{{\rm i}}
\def\rd{{\rm d}}

\def\re{{\rm e}}

\def\PT{$\mathcal{PT}$}

\begin{document}
\title{Nonlinear light behaviors near phase transition in non-parity-time-symmetric complex waveguides}
\author{Sean Nixon and Jianke Yang}
\affiliation{Department of Mathematics and Statistics, University of
Vermont, Burlington, VT 05401, USA}

\begin{abstract}
Many classes of non-parity-time (\PT) symmetric waveguides with
arbitrary gain and loss distributions still possess all-real linear
spectrum or exhibit phase transition. In this article, nonlinear
light behaviors in these complex waveguides are probed analytically
near a phase transition. Using multi-scale perturbation methods, a
nonlinear ordinary differential equation (ODE) is derived for the
light's amplitude evolution. This ODE predicts that the first class
of these non-\PT-symmetric waveguides support continuous families of
solitons and robust amplitude-oscillating solutions both above and
below phase transition, in close analogy with \PT-symmetric systems.
For the other classes of waveguides, the light's intensity always
amplifies under the effect of nonlinearity even if the waveguide is
below phase transition. These analytical predictions are confirmed
by direct computations of the full system.
\end{abstract}
\maketitle

The study of complex optical potentials which still possess an
all-real spectrum has its roots in parity-time (\PT) symmetry. First
introduced as a non-Hermitian generalization of quantum mechanics
\cite{Bender1998}, this concept later spread to optics, where an
even refractive index profile together with an odd gain and loss
landscape constitutes a \PT-symmetric system \cite{Musslimani2008}.
In this optical setting, \PT symmetry has been experimentally
realized \cite{Ruter_2010, Regensburger_2012, Peng_2014}. A
distinctive feature of \PT-symmetric systems is phase transition,
where the spectrum turns from all-real to partially-complex when the
gain-loss component (relative to the real refractive index) rises
above a certain threshold
\cite{Bender1998,Musslimani2008,Ruter_2010, Regensburger_2012,
Peng_2014,Ahmed2001}. This phase transition has been utilized in a
number of emerging applications of \PT optics
\cite{Feng2013,PTlaser_Zhang, PTlaser_CREOL}. The added effects of
nonlinearity into \PT-symmetric systems induce further novel
properties which are being actively explored \cite{Driben2011,
Zezyulin2012a, Nixon2012b, Segev2013, Kivshar2013, Kartashov2014,
Wimmer2015,Kivshar_review,Yang_review}.

The downside of \PT optics lies in its restrictive construction,
where the refractive index must be even while the gain and loss
profile must be odd. Relaxing this restriction, non-\PT-symmetric
complex potentials with all-real spectra have been introduced
\cite{Cannata1998,SUSY2013,Tsoy2014,PRA2016}. In particular, many
classes of non-\PT-symmetric complex potentials with arbitrary
gain-loss distributions and all-real spectra were reported in
\cite{PRA2016}. By tuning the free parameters in those complex
potentials, phase transition could also be induced. In the linear
regime, light in those different classes of non-\PT-symmetric
potentials behaves similarly \cite{PRA2016}. In the presence of
nonlinearity, will light still behave similarly in those different
classes of potentials?

In this article, we analytically probe nonlinear light behaviors in
these different classes of non-\PT-symmetric complex potentials with
all-real spectra and phase transition. Our analysis is focused near
a phase transition, where a pair of real eigenvalues of the
waveguide coalesce and form an exceptional point. Using multi-scale
perturbation methods, a nonlinear ordinary differential equation
(ODE) is derived for the light's amplitude evolution. This ODE
predicts that the first class of these non-\PT-symmetric waveguides
support continuous families of solitons and robust
amplitude-oscillating solutions both above and below phase
transition, closely resembling \PT-symmetric systems. For the other
classes of waveguides, the light's intensity always amplifies under
the effect of nonlinearity (even if the waveguide is below phase
transition). These analytical results show that different classes of
non-\PT-symmetric waveguides exhibit different nonlinear behaviors.

The paraxial model for the propagation of light down a waveguide
with a longitudinally uniform refractive index distribution and
gain-loss landscape is the potential nonlinear Schr\"{o}dinger (NLS)
equation
\begin{equation}
\ri \Psi_z + \Psi_{xx} +V(x) \Psi + \sigma |\Psi|^2 \Psi = 0,
\label{Eq:LatticeNLS}
\end{equation}
where $V(x)$ is a complex potential whose real part represents the
linear refractive index and whose imaginary part represents gain and
loss. If this potential satisfies the symmetry relation
\begin{equation} \label{Eq:Similarity}
\eta L = L^\dagger \eta,
\end{equation}
where $L = \partial_{xx} + V(x)$, $\dagger$ represents the adjoint
of an operator, and $\eta$ is a Hermitian operator (i.e.,
$\eta^\dagger=\eta$), then when $\eta$ has an empty kernel, the
eigenvalues of $L$ come in complex-conjugate pairs \cite{PRA2016}.
In that case, the spectrum of $L$ is either all-real, or partially
complex with conjugate pairs of eigenvalues. As an example, the
class of \PT-symmetric potentials, where $V^*(x)=V(-x)$, falls
within this framework when $\eta$ is chosen as the parity operator,
$\mathcal{P}f(x) \equiv f(-x)$. Here, $*$ represents complex
conjugation.

However, when $\eta$ is chosen as a differential operator, wide
classes of non-\PT-symmetric potentials with all-real spectra can be
obtained \cite{PRA2016}. One such class of potentials, which we
denote as Type-I, are
\begin{equation}\label{Eq:TypeI}
V(x; \gamma) = g^2(x) - 2\gamma g(x) + \ri g'(x),
\end{equation}
where $g(x)$ is an arbitrary real function, and $\gamma$ is a free
real parameter. For this type of potentials, the associated
Hermitian operator $\eta$ is
\begin{equation} \label{eta_typeI}
\eta(x; \gamma)=\ri \partial_x -g(x)+\gamma.
\end{equation}
Another class of such potentials, which we denote as Type-II, are
\begin{equation}\label{Eq:TypeII}
V(x; \gamma) = \frac{1}{4} (g^2-g_{\infty}^2)+ \frac{g'^2- 2g g''}{4g^2}  + \gamma \left( \frac{1}{ g^2 } - \frac{1}{g_{\infty}^2}\right)+ \ri g',
\end{equation}
where $g(x)$ is again an arbitrary real function, $g_{\infty} =
\lim_{x\rightarrow \pm \infty} g(x)$, and $\gamma$ is a free real
parameter. Here we have assumed that $g(x)$ approaches the same
non-zero constant as $x\to \pm \infty$ so that $V(x)$ is localized.

In both types of potentials, since $g(x)$ is arbitrary, the gain and
loss profile (i.e., the imaginary part) of the potential can be
arbitrary. Thus, these potentials are non-\PT-symmetric in general.
Yet, their spectra can still be all-real, which is surprising. Also
notice that both potentials contain a free parameter ($\gamma$). By
tuning this parameter, phase transition can be induced
\cite{PRA2016}.

\begin{figure}[!htbp]
    \centering
    \includegraphics[width=0.45\textwidth]{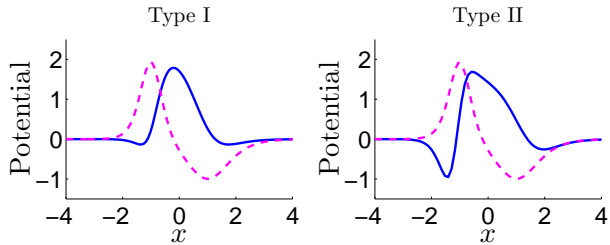}
    \caption{(Left) Type-I potential with $g(x)$ given by (\ref{g_typeI}) and
    $\gamma = 0.368$.
    (Right) Type-II potential with $g(x)$ given by (\ref{g_typeII}) and $\gamma = 0.467$. Solid blue lines are Re($V$) and
    dashed magenta Im($V$). }
    \label{Fig:Types}
\end{figure}

To illustrate, we take
\begin{equation}  \label{g_typeI}
g(x) = \tanh 2(x+1) - \tanh (x-1)
\end{equation}
in the Type-I potential (\ref{Eq:TypeI}), and
\begin{equation}  \label{g_typeII}
g(x) = \tanh 2(x+1) - \tanh (x-1)+1
\end{equation}
in the Type-II potential (\ref{Eq:TypeII}). These two potentials are
plotted in Fig.~\ref{Fig:Types}, for $\gamma = 0.368$ and $\gamma =
0.467$ respectively. They have the same gain-loss profile but
different refractive index distributions. Both potentials are
non-\PT-symmetric; yet, their spectra are all-real (which we have
verified numerically).

\begin{figure}[!htbp]
    \centering
    \includegraphics[width=0.45\textwidth]{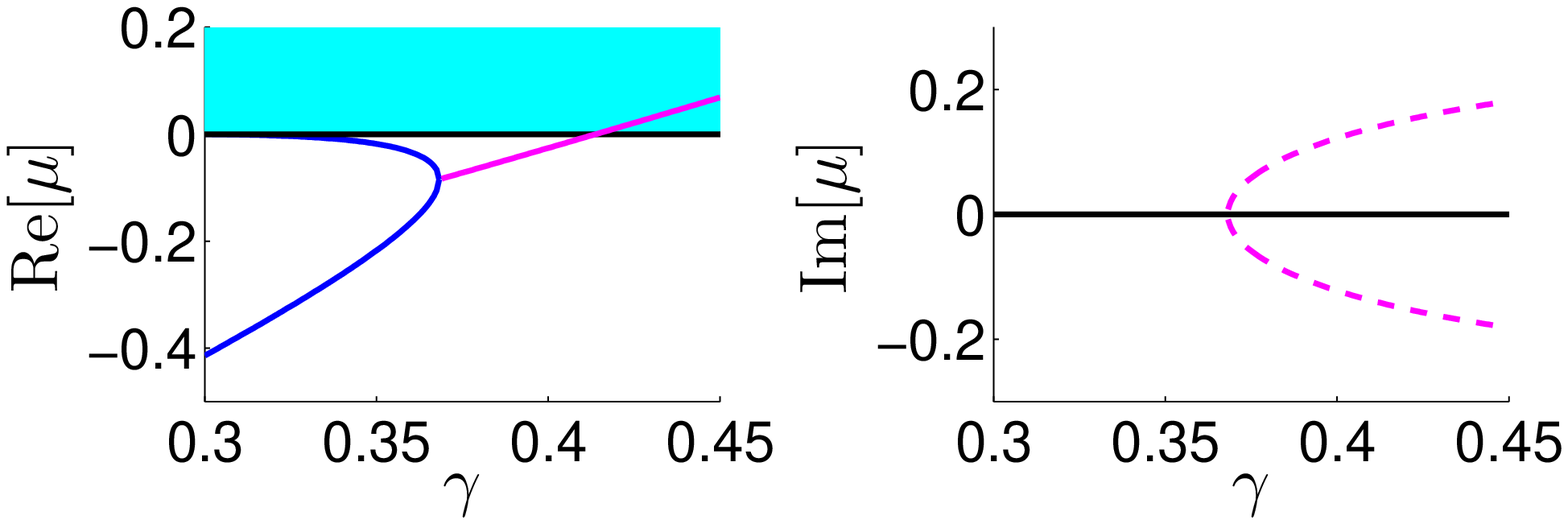}

  \vspace{0.2cm}
  \includegraphics[width=0.45\textwidth]{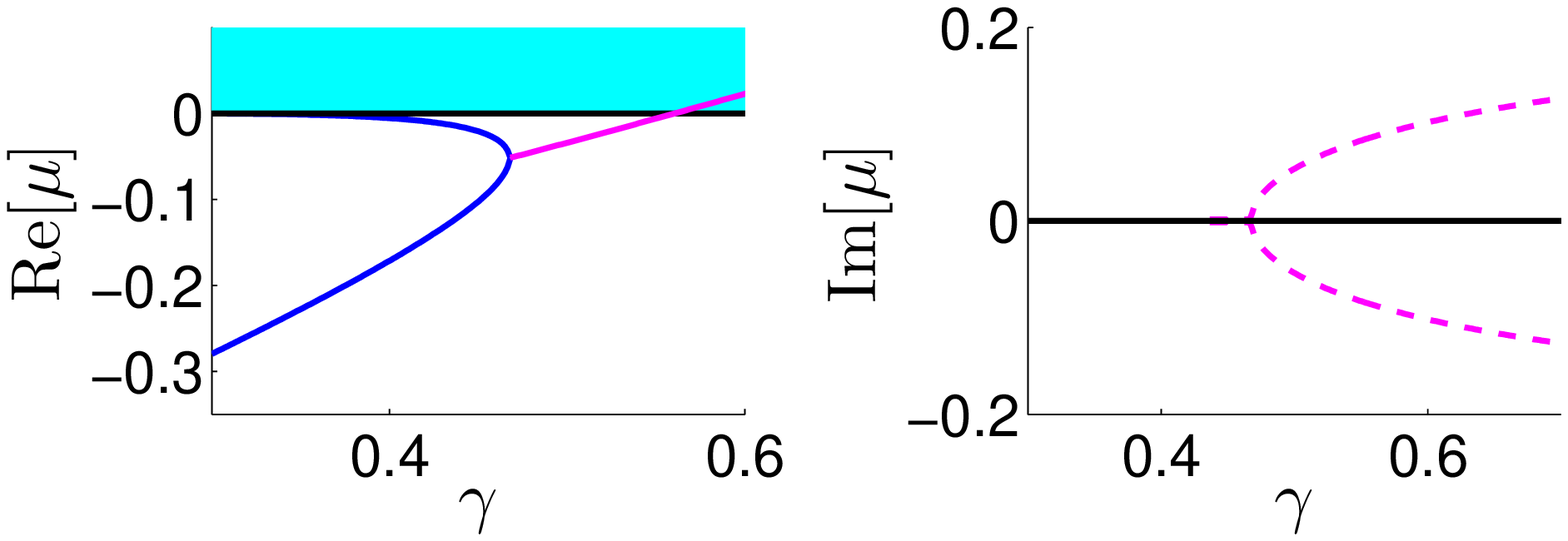}
    \caption{Phase transitions in Type-I (top) and Type-II (bottom) potentials.
     Real eigenvalues are in solid blue, complex eigenvalues in magenta
     and the cyan region representing the continuous spectrum.}
    \label{Fig:Spec}
\end{figure}

For the above $g(x)$ functions, varying the parameter $\gamma$ can
induce a phase transition. For example, for Type-I potentials, the
eigenvalue spectra for a range of $\gamma$ values are displayed in
Fig.~\ref{Fig:Spec} (top panel). Here, eigenvalues $\mu$ are defined
by $L\psi =-\mu \psi$. It is seen that at $\gamma \approx 0.368$,
two discrete real eigenvalues collide, form an exceptional point,
and then bifurcate off the real axis, thus a phase transition
occurs. The Type-I potential displayed in Fig.~\ref{Fig:Types} is at
this phase-transition point.

Similarly, the tuning of $\gamma$ can induce a phase transition in
the Type-II potentials. This is shown in Fig.~\ref{Fig:Spec} (bottom
panel), where a phase transition occurs at $\gamma \approx 0.467$.
The potential at this point was displayed in Fig.~\ref{Fig:Types}
(right panel).

It is easy to notice from Fig. \ref{Fig:Spec} that the spectra of
Type-I and Type-II potentials behave quite similarly, which
indicates that in the linear regime, solution dynamics in the two
types of potentials would be largely similar. Then, in the nonlinear
regime, would their solution dynamics still be similar? The answer
is negative, as we will show analytically below.

For our analytical treatment, we focus on these potentials near a
phase transition, where nonlinear solution dynamics can be
determined by multi-scale perturbation methods. Suppose the spectrum
of $L$ has an exceptional point at $\mu=\mu_0$ when $\gamma =
\gamma_0$ and consider the potentials nearby in the family, i.e,
$\gamma = \gamma_0 + \epsilon$, where $|\epsilon| \ll 1$. Then, up
to order $\epsilon$ the potential is
\begin{equation*}
V(x; \gamma) \approx V_0(x) + \epsilon V_2(x),
\end{equation*}
where $V_0(x)= V(x; \gamma_0)$ is the unperturbed potential at a
phase transition, and $V_2(x) = \frac{\partial }{\partial \gamma}
V(x; \gamma)|_{\gamma=\gamma_0}$. Defining
\begin{equation}
L_0 =  \partial_{xx}  + V_0(x) +\mu_0,
\end{equation}
then since $\mu_0$ is an exceptional point, $L_0$ has one
eigenfunction and at least one generalized eigenfunction where
\begin{equation}
\label{Eq:L0ueug}
L_0u_e=0, \quad L_0u_g=u_e.
\end{equation}
For simplicity, we assume the algebraic multiplicity of the
exceptional point $\mu_0$ is two, which is the generic case when two
simple real eigenvalues coalesce (see Fig. \ref{Fig:Spec}). Then,
$L_0$ has a single generalized eigenfunction, i.e., the equation
$L_0u_{g2}=u_g$ does not admit a localized solution.

When the symmetry relation (\ref{Eq:Similarity}) is evaluated at
$\gamma=\gamma_0$, we get $\eta_0 L_0 = L_0^\dagger \eta_0$, where
$\eta_0(x)= \eta(x;\gamma_0)$. From this, we see that $L_0^\dagger
\eta_0 u_e=0$, i.e., $\eta_0 u_e$ is in the kernel of the adjoint
operator $L_0^\dagger$. Thus, the solvability condition for the
$L_0u_g=u_e$ equation is that
\begin{equation}
\langle u_e, \eta_0 u_e \rangle = 0,
\end{equation}
where $\langle f, g \rangle = \int_{-\infty}^{\infty} f g^* \rd x $
is the standard inner product. In addition, the non-solvability
condition for the $L_0u_{g2}=u_g$ equation is that
\begin{equation}
D\equiv \langle u_g, \eta_0 u_e \rangle \ne 0.
\end{equation}
Note that from $L_0u_e=0$, we have $L_0^*u_e^*=0$. But for $L_0$,
$L_0^\dagger=L_0^*$, thus $L_0^\dagger u_e^*=0$. Recalling
$L_0^\dagger \eta_0 u_e=0$ above, we see that $\eta_0 u_e=Cu_e^*$,
where $C$ is some constant.

Now we consider low-amplitude nonlinear solutions to
Eq.~(\ref{Eq:LatticeNLS}) near this exceptional point. These
solutions can be expanded into a multi-scale perturbation series,
\begin{equation*}
\Psi(x, z) = \left( |\epsilon|^{\frac{1}{2}}\, u_1(x, Z) + |\epsilon| \hspace{0.04cm} u_2 +
|\epsilon|^{\frac{3}{2}} u_3 +\ldots\right)\re^{-\ri \mu_0 z},
\end{equation*}
where $Z = |\epsilon|^{\frac{1}{2}}\hspace{0.01cm} z$ is the
slow-modulation scale. At the first three orders, we obtain the
system of equations
\begin{align*}
L_0 u_1 & = 0, \hspace{1.1cm}  L_0 u_2 = -\ri u_{1Z},\\
L_0 u_3 & = -\ri u_{2Z} - \rho V_2 u_1 - \sigma |u_1|^2 u_1,
\end{align*}
where $\rho=\mbox{sgn}(\epsilon)$. The $u_1$ and $u_2$ equations can
be solved:
\begin{equation*}
u_1= A(Z) u_e(x), \quad
u_2 = -\ri A'(Z) u_g(x).
\end{equation*}
Substituting them into the $u_3$ equation, we get
\begin{equation*}
L_0 u_3 = - A_{ZZ} u_g - A \rho V_2 u_e - \sigma |A|^2 A |u_e|^2 u_e,
\end{equation*}
whose solvability condition yields
\begin{equation}
A_{ZZ} - \alpha A + \sigma_1 |A|^2 A = 0,
\label{Eq:AFull}
\end{equation}
where
\begin{equation}
\alpha =  - \frac{\rho}{D}   \langle V_2 u_e, \eta_0 u_e \rangle, \quad
\sigma_1 =  \frac{\sigma}{D} \langle  |u_e|^2 u_e, \eta_0 u_e  \rangle.
\end{equation}
Eq. (\ref{Eq:AFull}) is our reduced ODE model for nonlinear solution
dynamics near an exceptional point in these non-\PT-symmetric
complex potentials.

Now we show that in this ODE model, $\alpha$ is always real. In
addition, $\sigma_1$ is real for Type-I potentials but complex for
other potentials such as Type-II.

First, we note that since $L_0$ satisfies the symmetry relation
$\eta_0 L_0 = L_0^\dagger \eta_0$, then
\begin{align*}
D^*&= \langle \eta_0 u_e , u_g  \rangle =\langle \eta_0 L_0 u_g , u_g  \rangle  = \langle L_0^\dagger \eta_0 u_g, u_g  \rangle \\
& = \langle  \eta_0 u_g, L_0 u_g  \rangle  = \langle  \eta_0 u_g, u_e  \rangle = \langle   u_g, \eta_0 u_e \rangle = D,
\end{align*}
thus $D$ is real.

Next, we differentiate the symmetry equation (\ref{Eq:Similarity})
with respect to the free parameter $\gamma$, which yields
$\eta_{\gamma 0}L_0+\eta_0 V_2 = V_2^* \eta_0+L_0^\dagger
\eta_{\gamma 0}$, where $\eta_{\gamma 0} = \frac{\partial }{\partial
\gamma} \eta(x; \gamma)|_{\gamma=\gamma_0}$. Then,
\begin{align*}
& \langle V_2 u_e, \eta_0 u_e \rangle^* = \langle \eta_0 u_e , V_2 u_e  \rangle =\langle V_2^* \eta_0 u_e , u_e  \rangle  \\
&= \langle  (\eta_0  V_2 +\eta_{\gamma 0}L_0-L_0^\dagger\eta_{\gamma 0} ) u_e, u_e  \rangle \\
& =\langle  \eta_0 V_2 u_e, u_e \rangle -\langle \eta_{\gamma 0} u_e, L_0u_e  \rangle = \langle  V_2 u_e, \eta_0 u_e \rangle,
\end{align*}
thus, $\alpha$ is real.

For the nonlinear coefficient $\sigma_1$, the symmetry relation
alone is insufficient to guarantee a real constant. Instead, the
form (\ref{eta_typeI}) of operator $\eta$ associated with Type-I
potentials must be employed. In this case,
\begin{align*}
\langle  |u_e|^2 u_e, \eta_0 u_e  \rangle &= \langle  |u_e|^2 u_e, (\ri \partial_x -g+\gamma_0) u_e  \rangle \\
& \hspace{-2.2cm} =   \frac{1}{2}\Imag \langle u_e^{2}, (u_e^2)_x
\rangle - \int_{-\infty}^{\infty} (g-\gamma_0) |u_e|^4\rd x,
\end{align*}
thus, $\sigma_1$ is real for Type-I potentials. However, for other
types of potentials, it is generically complex as can be numerically
verified.

The fact that $\sigma_1$ is real for Type-I potentials but complex
for other types of potentials has far-reaching consequences on the
nonlinear dynamics in these potentials. When $\sigma_1$ is real, the
ODE (\ref{Eq:AFull}) admits solutions with stationary amplitude,
$A(Z) = A_0 \re^{- \ri \mu_1 Z}$, where $\mu_1$ is a free real
parameter, and $A_0$ is a constant given by
\begin{equation}\label{Eq:ASoliton}
|A_0|  =  \sqrt{(\mu_1^2  + \alpha)/\sigma_1}.
\end{equation}
This implies that under Type-I potentials, the full system
(\ref{Eq:LatticeNLS}) admits continuous families of solitons,
$\Psi(x; \mu) = u(x) \re^{-\ri \mu z}$, parameterized by $\mu$,
where
\[
\mu=\mu_0 + |\epsilon|^{\frac{1}{2}} \mu_1, \quad
u(x)\approx |\epsilon|^{\frac{1}{2}}A_0u_e(x).
\]
In particular, when $\alpha>0$ (above phase transition) and
$\sigma_1>0$, Eq.~(\ref{Eq:ASoliton}) predicts a soliton family
above an amplitude threshold $|A_0|_{min} = \sqrt{\alpha/\sigma_1}$;
when $\alpha<0$ (below phase transition) and $\sigma_1<0$,
Eq.~(\ref{Eq:ASoliton}) predicts a soliton family below an amplitude
threshold $|A_0|_{max}=\sqrt{\alpha/\sigma_1}$. These predictions
are verified in our numerics of the full system
(\ref{Eq:LatticeNLS}) (see also
\cite{Tsoy2014,Konotop_OL2014,SAPPM2015}).

However, when $\sigma_1$ is complex, the ODE (\ref{Eq:AFull}) does
not admit solutions with stationary amplitude. This implies that
under Type-II and higher-type potentials, the full system
(\ref{Eq:LatticeNLS}) does not admit continuous families of
solitons. This is consistent with our direct numerics on this
system, where soliton solutions could not be found.

\begin{figure}[!htbp]
    \centering
    \includegraphics[width=0.45\textwidth]{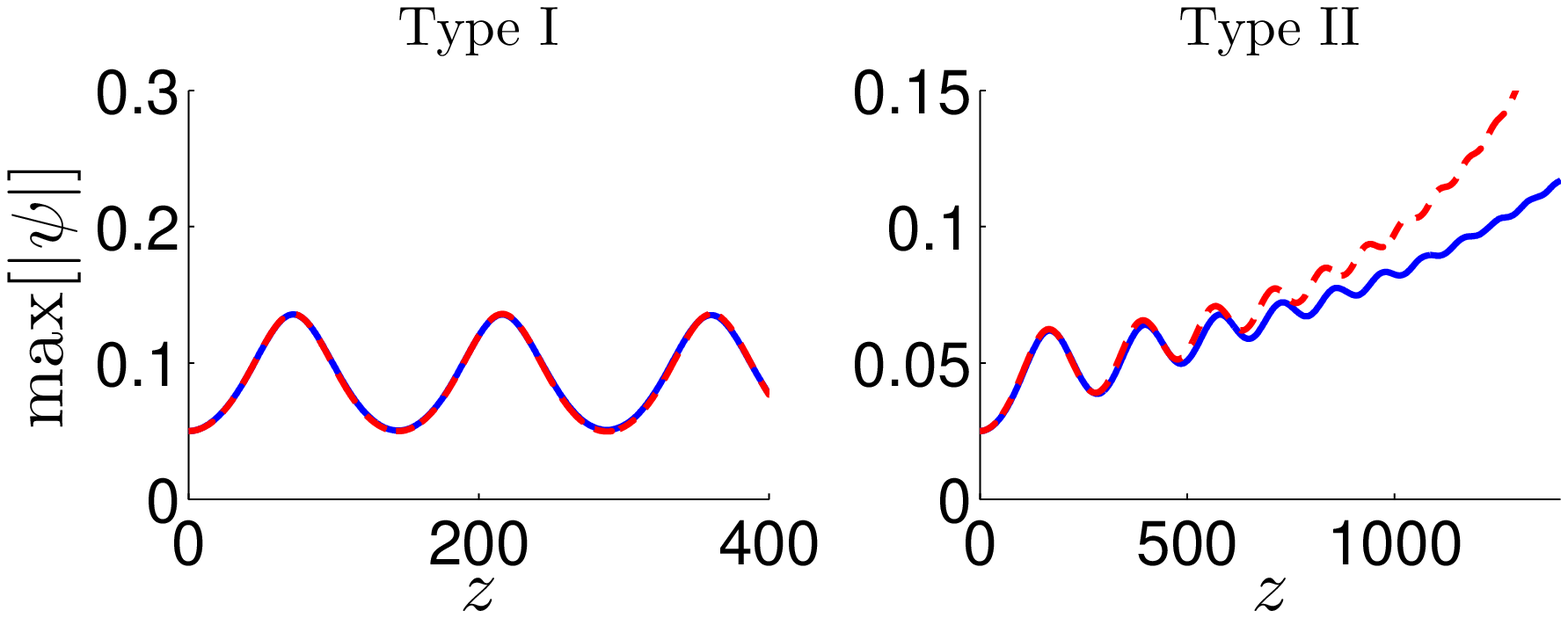}

    \vspace{0.2cm}
   \includegraphics[width=0.45\textwidth]{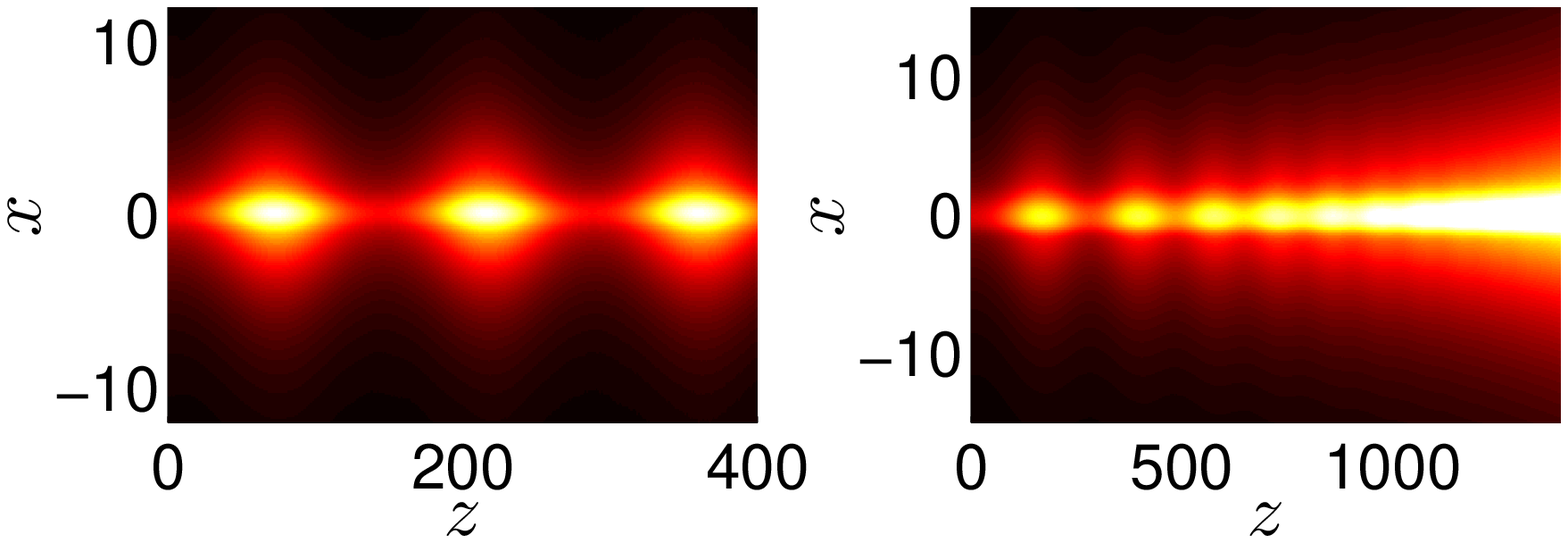}
    \caption{Top: Comparisons of numerical simulations of the full system in solid blue against ODE results in
    dashed red for potentials above phase transition with focusing nonlinearity.
    Bottom: Solution evolutions in the full system. Left: Type-I potential. Right: Type-II potential. }
    \label{Fig:Above}
\end{figure}

Beyond solitons, the ODE (\ref{Eq:AFull}) also predicts different
behaviors on other types of solutions for real and complex
$\sigma_1$ values. If $\sigma_1$ is real, Eq. (\ref{Eq:AFull})
exhibits periodic orbits when $\sigma_1>0$ (for either sign of
$\alpha$). This implies that under Type-I potentials, the full
system (\ref{Eq:LatticeNLS}) admits periodically-oscillating
solutions under a suitable sign of nonlinearity both above and below
phase transition. For example, for the Type-I potential with $g(x)$
given by (\ref{g_typeI}) and $\epsilon=0.05^2$ (above phase
transition), a periodic orbit of the ODE under focusing nonlinearity
($\sigma=1$) is plotted in Fig. \ref{Fig:Above} (upper left panel).
Numerically, we have found the corresponding solution in the PDE
system (\ref{Eq:LatticeNLS}), which is shown in the left panels of
the same figure.

\begin{figure}[!htbp]
    \centering
    \includegraphics[width=0.45\textwidth]{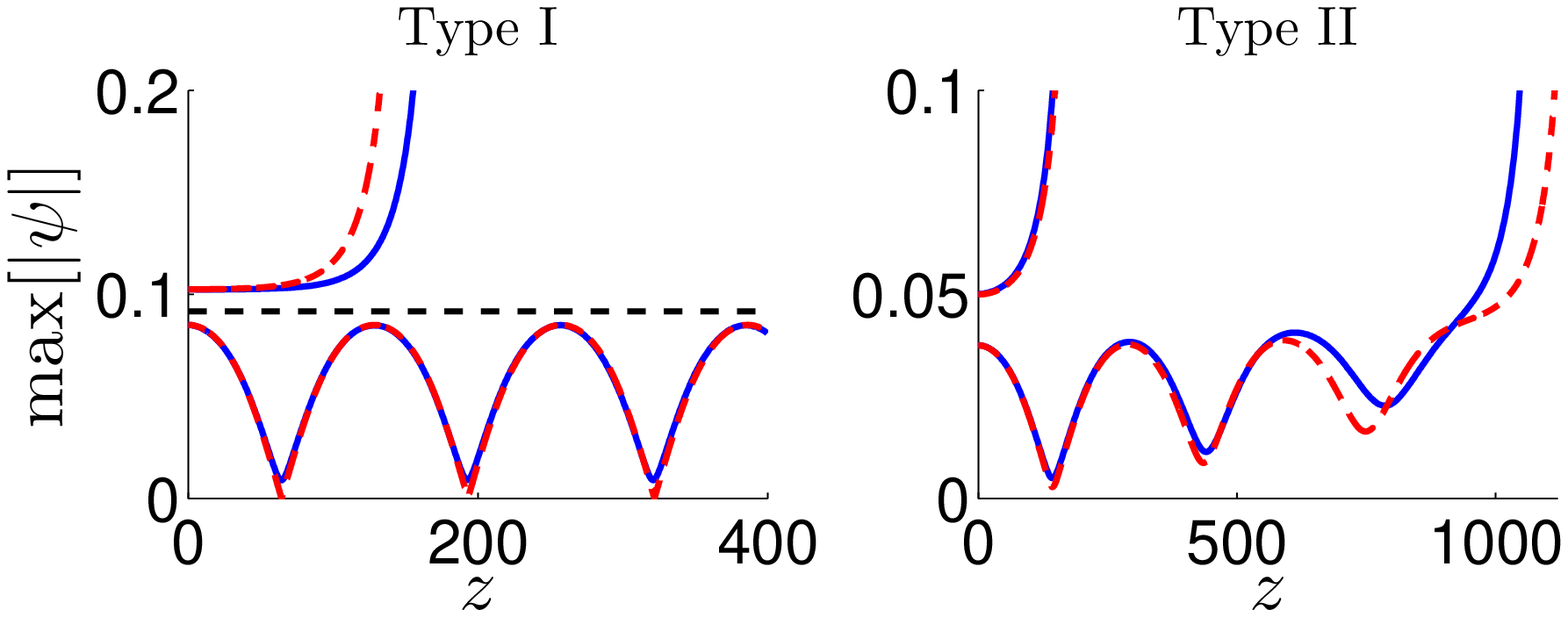}

    \vspace{0.2cm}
   \includegraphics[width=0.45\textwidth]{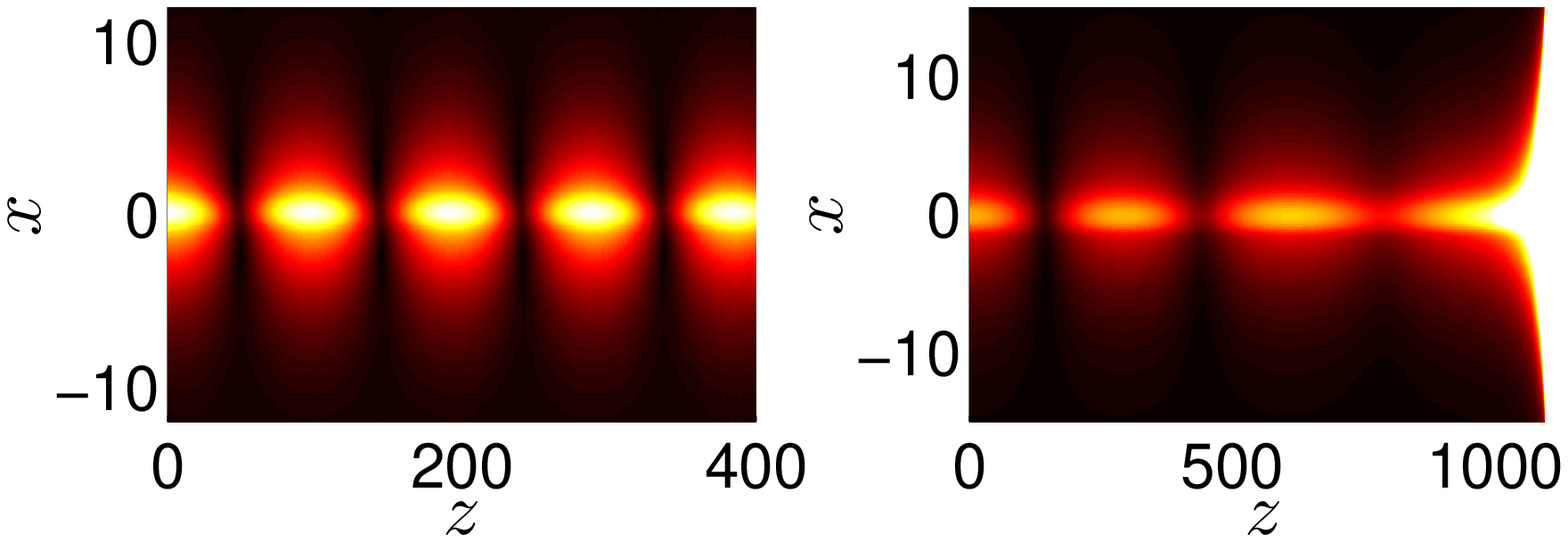}
    \caption{Top: Comparisons of numerical simulations of the full system in solid blue against ODE results in
    dashed red for potentials below phase transition with defocusing nonlinearity.
    Bottom: Solution evolutions in the full system for the lower ODE orbits in the top panels. }
    \label{Fig:Below}
\end{figure}

If $\sigma_1$ is real, Eq. (\ref{Eq:AFull}) also exhibits periodic
orbits when $\sigma_1<0$, $\alpha<0$, and the initial amplitude is
below a certain threshold (above this threshold, the ODE solution
will blow up to infinity in finite distance). This implies that
under Type-I potentials, the full system (\ref{Eq:LatticeNLS})
admits periodically-oscillating solutions (at low amplitude) and
amplifying solutions (at high amplitude) under an opposite sign of
nonlinearity below phase transition. As an example, we consider the
same Type-I potential but with defocusing nonlinearity and
$\epsilon=-0.05^2$ (below phase transition). In this case, two ODE
orbits, one periodic and the other amplifying, are plotted in
Fig.~\ref{Fig:Below} (upper left panel). These orbits match the
corresponding solutions in the PDE system shown in the left panels
of the same figure.

However, if $\sigma_1$ is complex, the ODE (\ref{Eq:AFull}) does not
admit periodic orbits. Instead, all solutions will amplify and blow
up to infinity in finite distance. This implies that for other
potentials such as Type-II, the PDE solutions will always amplify to
high amplitude regardless of the sign of nonlinearity or whether the
potential is above or below phase transition. To verify this
behavior, we first consider an example Type-II potential with $g(x)$
given by (\ref{g_typeII}), where a phase transition occurs at
$\gamma_0\approx 0.467$ and $\mu_0\approx -0.052$. In this case, the
coefficients in the ODE model are $\alpha = {\rm
sgn}(\epsilon)0.093$ and $\sigma_1=\sigma (0.104+ \ri 0.011)$, where
$u_e(x)$ has been normalized to have unit peak amplitude. For
$\epsilon=0.05^2$ (above phase transition) and $\sigma=1$ (focusing
nonlinearity), an amplifying solution in the ODE model and the
corresponding PDE solution are displayed in the right panels of Fig.
\ref{Fig:Above}, where good agreement can be seen. To verify the
amplifying behavior below phase transition, we take this same
Type-II potential but with $\epsilon=-0.05^2$ and defocusing
nonlinearity. In this case, an amplifying ODE solution and the
corresponding PDE solution are displayed in the right panels of Fig.
\ref{Fig:Below}.

The above nonlinear dynamics in Type-I potentials is remarkably
similar to that in \PT-symmetric potentials \cite{PhysicaD2016}. In
contrast, nonlinear dynamics in Type-II and higher-type potentials
is quite different. Thus, we conclude that, even though all
non-\PT-symmetric potentials obeying the symmetry relation
(\ref{Eq:Similarity}) for differential operators $\eta$ exhibit
similar linear behaviors, there is a large difference between Type-I
potentials and the other potentials on nonlinear behaviors.

Why are Type-I potentials so special on nonlinear dynamics? One
reason is that only for these potentials does the full system
(\ref{Eq:LatticeNLS}) admit a conservation law,
\begin{equation}
\frac{\partial }{\partial z}\left[\Psi (\eta \Psi)^*\right]+\frac{\partial J}{\partial x}
=0,
\end{equation}
where $\eta$ is given in Eq. (\ref{eta_typeI}), and the flux
function $J$ is
\[
J=\Psi\Psi_{xx}^*-|\Psi_x|^2+\frac{1}{2}\sigma |\Psi|^4-i(g-\gamma)(\Psi\Psi_x^*-\Psi_x\Psi^*)-ig'|\Psi|^2.
\]
The associated conserved quantity is $I=\langle \Psi, \eta \Psi
\rangle$, where $\frac{dI}{dz}=0$. For the other types of
potentials, we could not find a conservation law. Apparently, this
conservation law puts strong restrictions on the nonlinear dynamics
in Type-I potentials.

In summary, nonlinear light behaviors in non-\PT-symmetric complex
waveguides obeying the symmetry condition (\ref{Eq:Similarity}) were
probed analytically near a phase transition. It was found that
nonlinear behaviors in Type-I waveguides are similar to those in
\PT-symmetric systems, while those in other waveguides behave quite
differently. Thus, these non-\PT-symmetric waveguides exhibit a
wider variety of nonlinear dynamics than \PT-symmetric waveguides.

This work was supported in part by Air Force Office of Scientific
Research (USAF 9550-12-1-0244) and National Science Foundation
(DMS-1311730).

\end{document}